\documentclass[a4paper]{jpconf}
\usepackage{graphicx}
\usepackage{textgreek}

\usepackage{lineno}

\begin{document}
\title{The Time Structure of Hadronic Showers in Calorimeters with Scintillator and with Gas Readout}

\author{Marco Szalay on behalf of the CALICE collaboration}

\address{Max Planck Institut f\"ur Physik,\\
F\"ohringer Ring 6, 80805 M\"unchen, Germany}

\ead{marco.szalay@mpp.mpg.de}

\begin{abstract}
Hadronic showers are characterized by a rich particle structure in the spatial as well as in the time domain. The prompt component comes from relativistic fragments that deposit energy at the ns scale, while late components are associated predominantly with neutrons in the cascade. To measure the impact of these late components, two experiments, based on gaseous and plastic active layers with steel and tungsten absorbers, were set up. The different choice for the material of the active layers produces distinct responses to neutrons, and consequently to late energy depositions. After discussing the technical aspects of these systems, we present a comparison of the signals, read out with fast digitizers with deep buffers, and provide detailed information of the time structure of hadronic showers over a long sampling window.
\end{abstract}

\section{Introduction} \label{section:intro}
The timing resolution of calorimeters depends strongly on the time structure of hadronic showers. The study of this quantity is of particular interest for precision signal time stamping that would allow to reject pile-up background from hadrons produced in two-photon processes at future accelerators such as the Compact Linear Collider (CLIC\cite{lebrun2012clic,linssen2012physics}). 

In a calorimeter, the time distribution of energy deposition depends on several physical processes taking place within the hadronic shower.
Relativistic hadrons and the electromagnetic fraction of the shower produce an intense prompt signal in the detector that extends for a few ns before fading out. After that, a rich set of nuclear interactions with the dense absorbing material releases neutrons that can be detected in the active layer via elastic scattering on hydrogen nuclei on a timescale that ranges over a few tens of ns. Even more delayed signals, up to ms, arise from energy released by neutron capture events and the decay of meta-stable nuclear states.

The sensitivity to neutron elastic scattering, as well as to the later parts of the hadronic cascade, also influences the spatial structure of the visible signal in the detector, especially when the absorbing material is tungsten, where the electromagnetic subshowers are denser, due to the shorter radiation length of the material.
The extent of the cascade is of particular importance for the performance of particle flow algorithms\cite{thomson2009particle,marshall2012performance}.

These algorithms, which are used for jet reconstruction with unprecedented precision in linear collider detectors, rely on two-particle separation in the calorimeters. The spatially resolved measurement of the time structure of showers and the comparison of detection media with different sensitivity to neutrons, such as plastic scintillators and gaseous detectors, is thus of high relevance for the development of calorimeter technologies for such a future collider.

\section{Experimental Setup}
\begin{figure}
  \centering
  \includegraphics[width=0.99\textwidth]{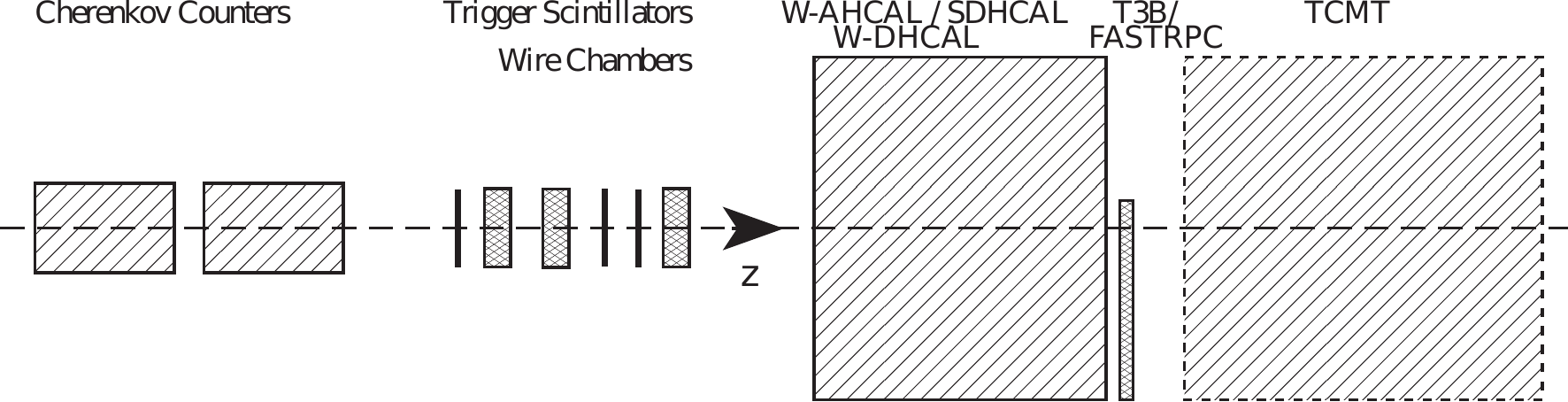}
    \caption{\label{fig:setup}Setup schematics at the CERN SPS H8 beam line in the North Hall. The tail catcher TCMT was only present for runs with the CALICE W-AHCAL and W-DHCAL, and not installed for runs with the Fe-SDHCAL. Illustration not to scale.}
\end{figure}
Two complementary experiments, based on different signal detection technologies, have been developed to take data together with the CALICE HCAL prototypes and measure the average time structure of hadronic showers, namely {\it T3B} (Tungsten Timing Test Beam\cite{simon2013t3b}) and {\it FastRPC}. The former consists of 15 3$\times$3 cm$^{2}$ BC420
scintillating tiles, each coupled to a Hamamatsu SiPM, arranged in a stripe to sample the hadronic shower radially from its center. The latter uses a glass RPC active layer, operated in avalanche mode, with readout pads that have the same geometry as T3B.

The setups share the same readout based on 15 (one per channel) Infineon SiGe preamplifiers\footnote{Infineon BGA614 (http://www.infineon.com/)} and four 4-channel USB-oscilloscopes\footnote{PicoTech PicoScope 6403 (http://www.picotech.com/)} which provide a sampling rate of 1.25 GS/s per channel with a buffer large enough to cover a 2.4 \textmu{}s acquisition window for each event. 
This allows the study of time structure of the energy deposits in the active medium in detail over a long time window, while maintaining a sub-ns resolution.

The small number of readout channels is insufficient for event-by-event measurements, but is used to measure the average time structure of showers in large data samples. To produce the results presented here, sizable data sets, summarized in table \ref{table:nEvents}, have been collected.
\begin{table}[h]
\centering
  \begin{tabular}{|l|c|}
    \hline
    Data Set & Number of Events \\ \hline \hline
    T3B 180 GeV \textmu{} & 5.40 M \\
    T3B 60 GeV hadrons (steel) & 1.60 M \\
    T3B 60 GeV hadrons (tungsten) & 4.06 M \\
    FastRPC 180 GeV \textmu{} & 3.19 M \\
    FastRPC 80 GeV hadrons (tungsten) & 2.63 M \\
    \hline
  \end{tabular}
  \caption{\label{table:nEvents}Statistics recorded for each data sample at CERN SPS test beam facility.}
\end{table}
T3B has been installed behind the tungsten analog scintillator (W-AHCAL\cite{adloff2010construction}) prototype as well as behind the steel semi-digital prototype (SDHCAL\cite{laktineh2011construction}). The FastRPC experiment was placed downstream to the tungsten digital RPC prototype (W-DHCAL\cite{bilki2008calibration}). All systems were installed at the CERN SPS facility and took data at beam energies up to 180 GeV. A schematic drawing of the test beam setup can be found in figure \ref{fig:setup}.
The material budget in the beam line upstream of the experiment's layer differs for tungsten and steel runs. The total amount of material is $\sim$ 5.1 nuclear interaction lengths for tungsten data and $\sim$ 6.5 $\lambda_I$ for steel.

The trigger system is shared with the HCAL prototypes and consists of two 10$\times$10 cm$^{2}$ scintillators in coincidence placed in front of the HCAL prototypes. 

\section{Results}
For both experiments, a sophisticated calibration and reconstruction framework has been developed. In the case of scintillator readout, this system is capable of determining the arrival time of each photon on the photon sensor on the nanosecond level by iteratively subtracting single photon signals from the recorded waveform\cite{simon2013t3b}. Further analysis is then performed on the photon time distribution. With RPC readout, the recorded pulse is used directly in the analysis.

To provide a robust base for comparison between the two systems and to eliminate effects from afterpulsing, the time of first hit is studied, which is defined by the time of the first energy deposit corresponding to at least the equivalent of 0.3 minimum-ionizing particles within 9.6 ns in a given cell in an event. Due to the high granularity of the readout, the probability for multiple hits in one cell in one event is on the percent level. The use of the time of first hit rather than using all observed hits thus does not result in an appreciable bias.
\begin{figure}
  \centering
  \includegraphics[width=0.49\textwidth]{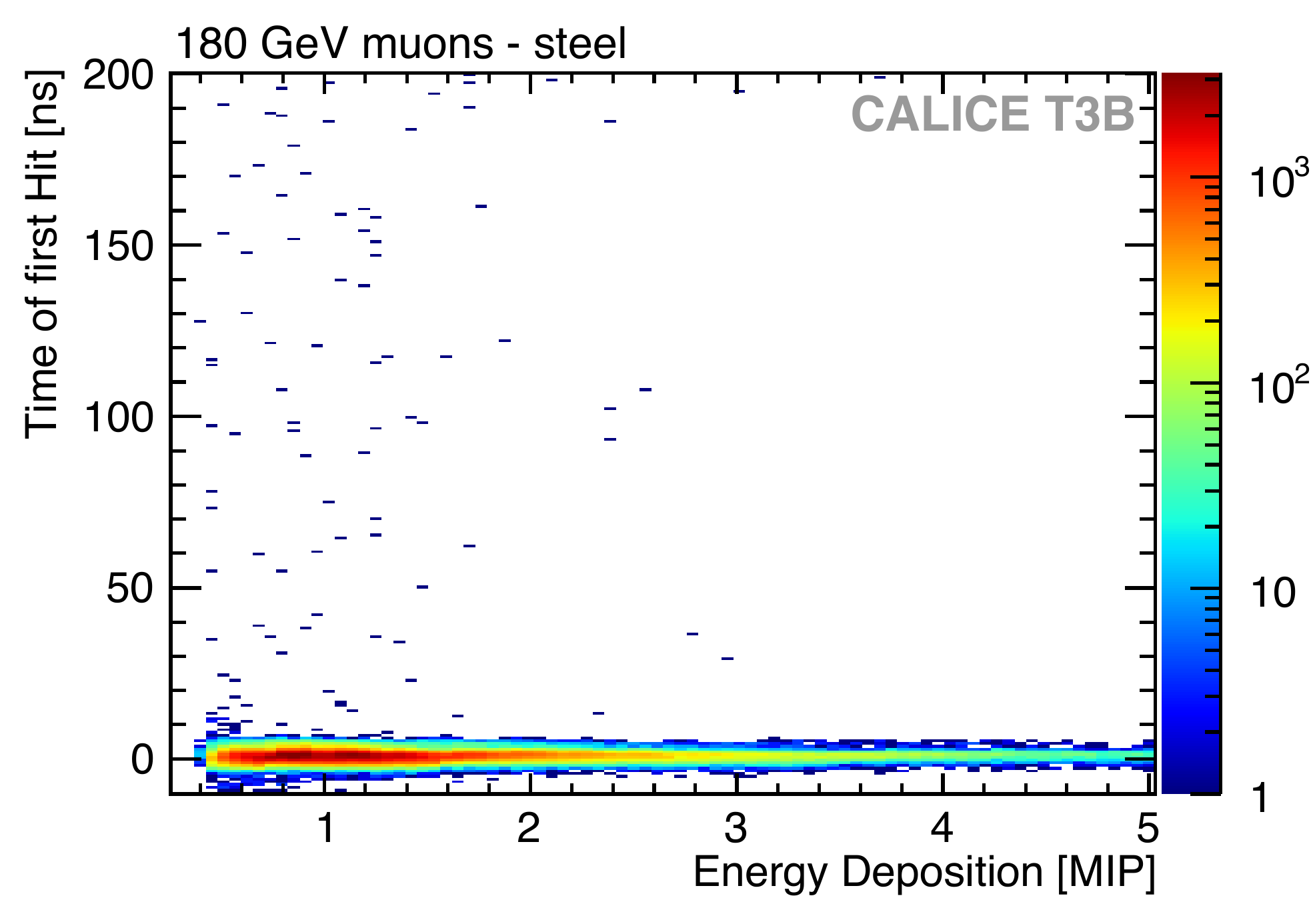}\\
  \includegraphics[width=0.49\textwidth]{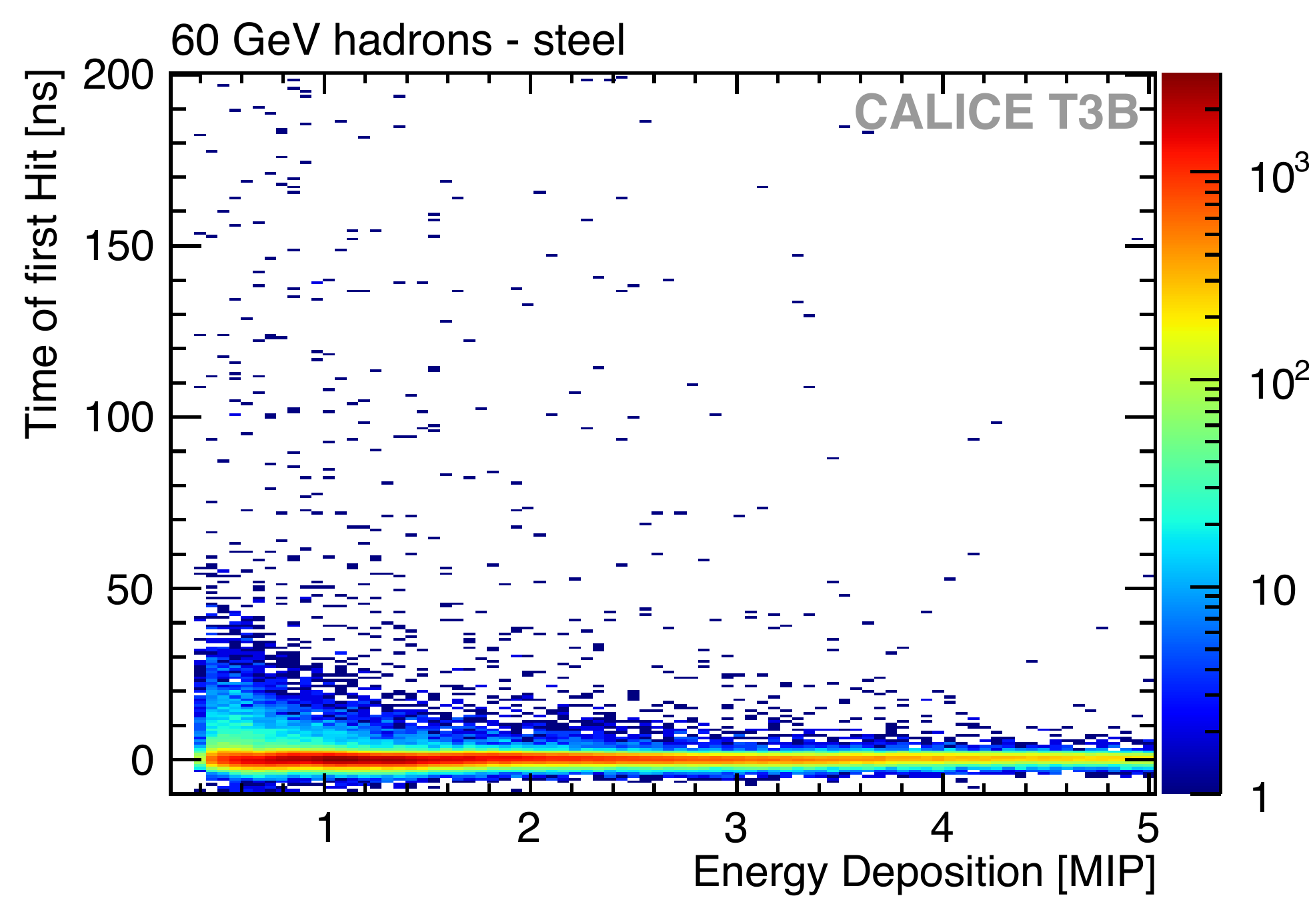}
  \includegraphics[width=0.49\textwidth]{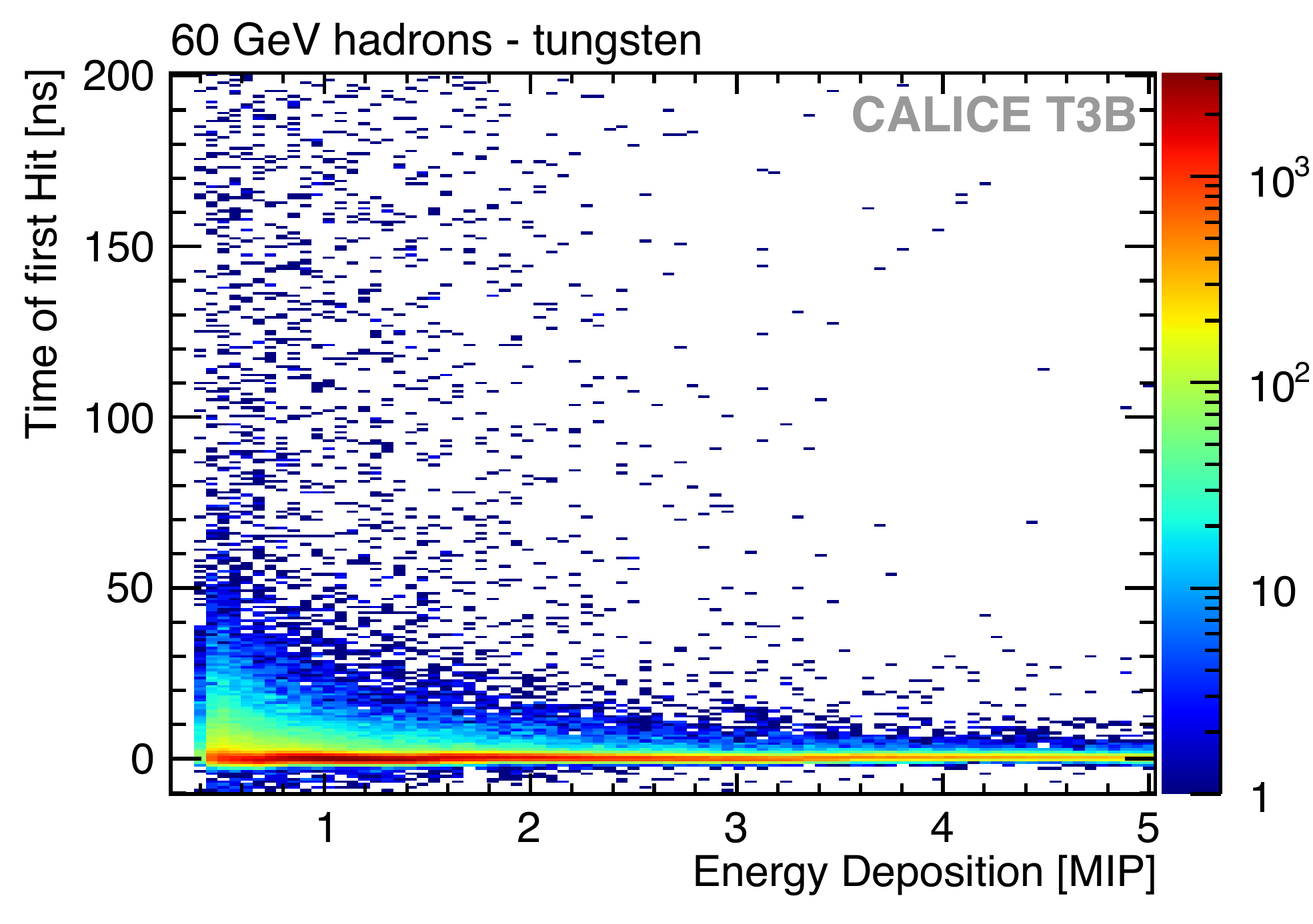}
  \caption{\label{fig:T3B_ToFH_vs_E}Distribution of reconstructed time of first hit in T3B as a function of deposited energy for muon data (top) and 60 GeV hadrons in steel (bottom, left) and in tungsten (bottom, right). With both absorbing materials, the time structure for hadrons extends over a longer time window, compared to muon data.}
\end{figure}

\begin{figure}
  \centering
  \includegraphics[width=1\textwidth]{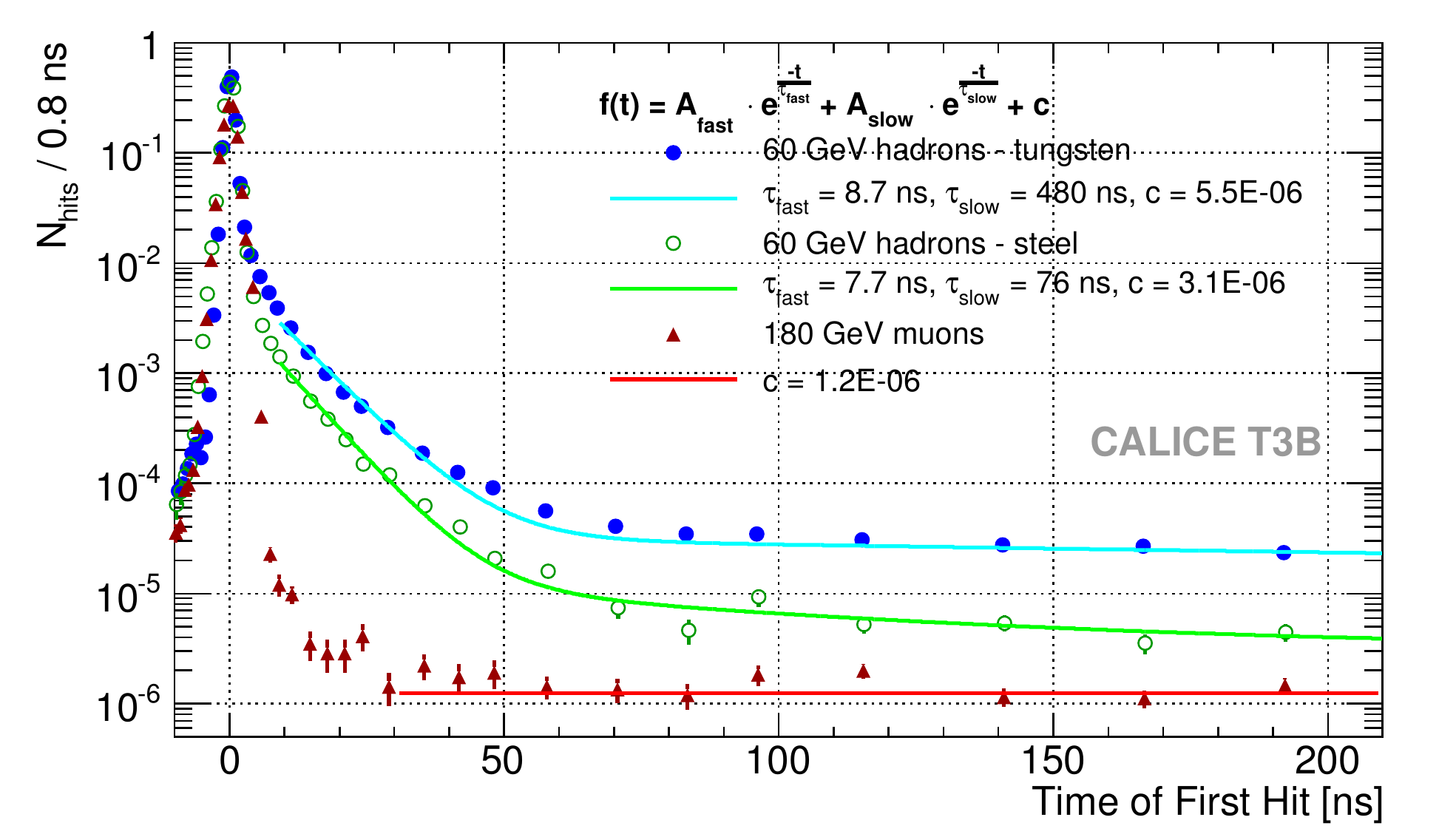} 
    \caption{\label{fig:TungstenvsSteel}Time of first hit distribution of muon data with steel absorbers and hadron data with steel and tungsten absorbers in a time range of -10 ns to 200 ns. The histograms are normalized to the number of events in which at least one first hit could be identified and show the number of hits per T3B DAQ time bin of 0.8 ns.}
\end{figure}

Figure \ref{fig:T3B_ToFH_vs_E} shows a comparison of the aforementioned time of first hit for T3B data\cite{adloff2014time}. Muon data show a prompt energy deposition compatible with relativistic particles depositing energy via ionization processes. In contrast, hadron data show a richer tail in the time domain due to slower components in the shower produced by particles interacting with the absorbing material. Moreover tungsten shows a higher non-prompt energy deposition activity compared to steel, a hint that absorber materials with a high neutron content release an increased number of evaporation neutron that account for a more prominent tail.
The discrepancy can be better appreciated in figure \ref{fig:TungstenvsSteel}\cite{adloff2014time}, where all T3B data sets are projected on the time axis.
Here the sum of two exponential decays and a noise constant term are used to fit the signal tail. The two exponentials account for a faster tail component, in the time region from a few ns up to $\sim50$ ns, that transitions smoothly into a slower shower component that extends to hundreds of ns. The time range of the first one is compatible with fast neutrons that scatter elastically on hydrogen nuclei present in the active material. The second one can be associated primarily with neutron capture, which release detectable energy over a long period of time.
\begin{figure}
  \centering
  \includegraphics[width=0.76\textwidth]{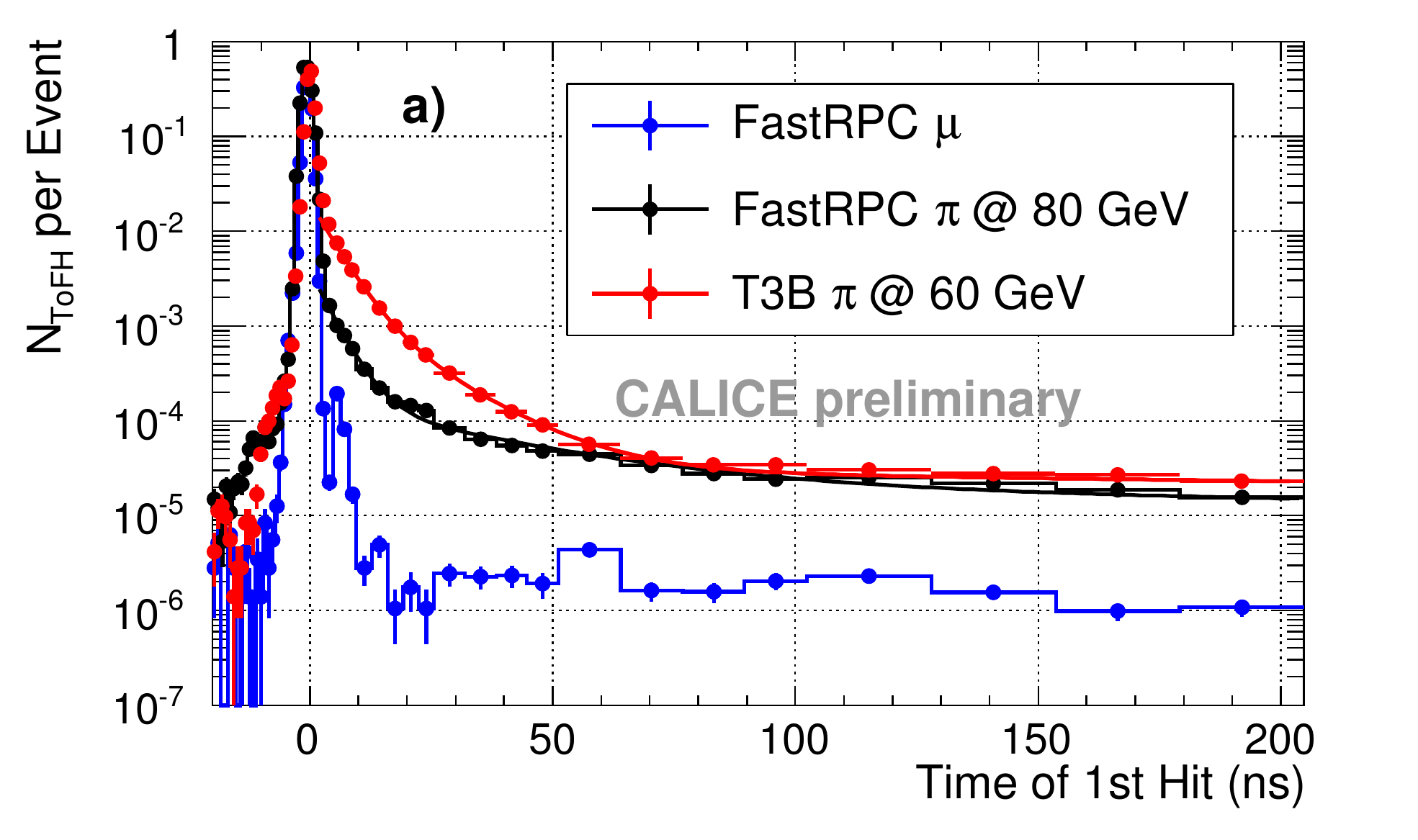}
  \includegraphics[width=0.76\textwidth]{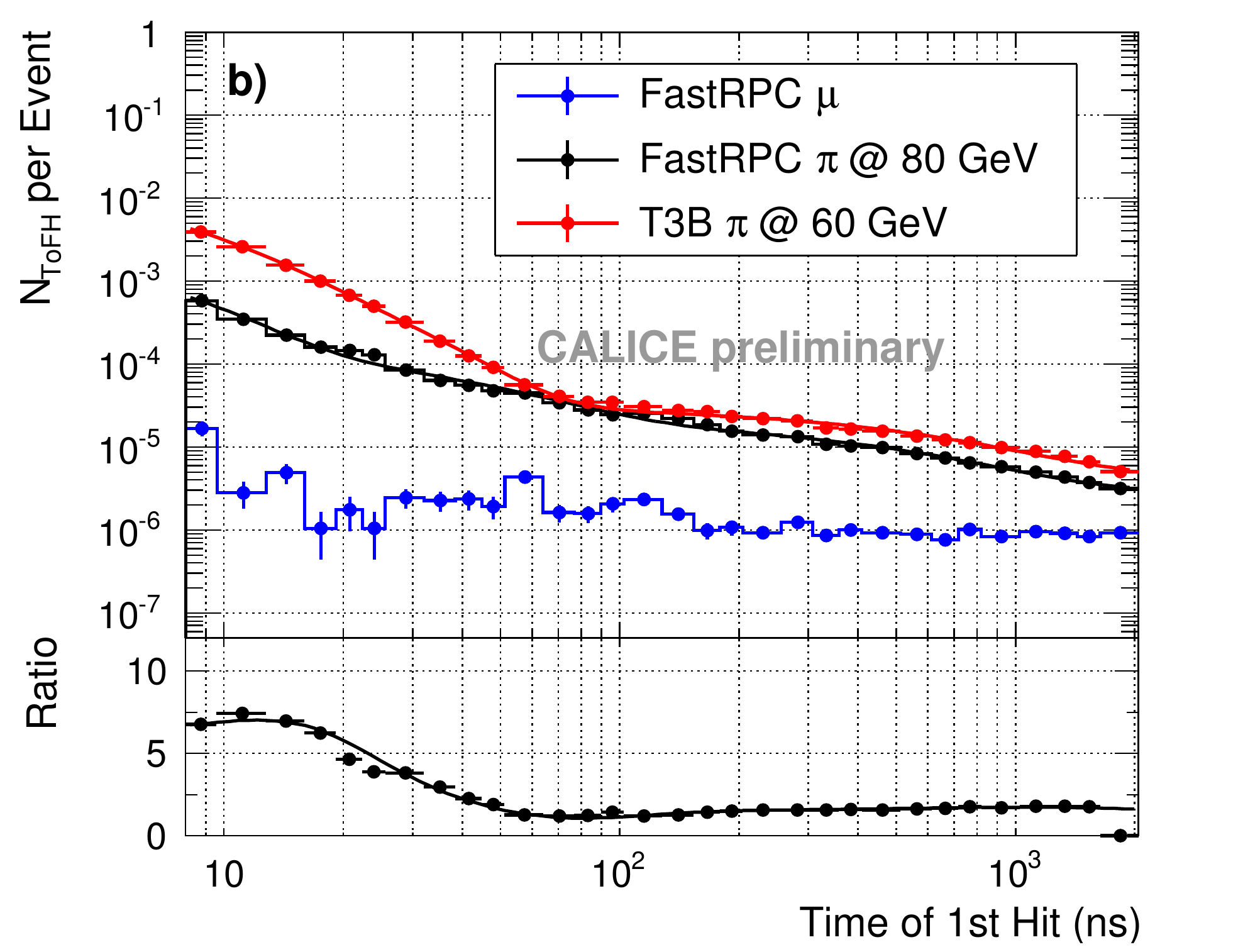}
  \caption{\label{fig:ScintRPC}Comparison of the time of first hit with scintillator and RPC readout in tungsten in the [-20, 200] ns range (top) and on a logarithmic time scale for [8, 2000] ns (bottom). For reference, the time distribution for muons in FastRPC is also shown.}
\end{figure}

A similar comparison can be performed, for tungsten absorbers, between the two different experiments, as depicted in figure \ref{fig:ScintRPC}. It is apparent that, whereas for both active materials hadronic showers lead to substantial late signal components, in the range from 10 to 50 ns there is a large difference, up to a factor eight, between the two setups. In this time window MeV-scale neutrons scattering elastically on hydrogen account for the largest fraction of the signal. Due to the low density and the low hydrogen content in the RPCs, their sensitivity to this component is significantly reduced compared to plastic scintillator. This discrepancy is then again strongly reduced for more delayed signals, where the contribution of neutron elastic scattering in the active layer is negligible.

\begin{figure}
  \centering
  \includegraphics[width=0.49\textwidth]{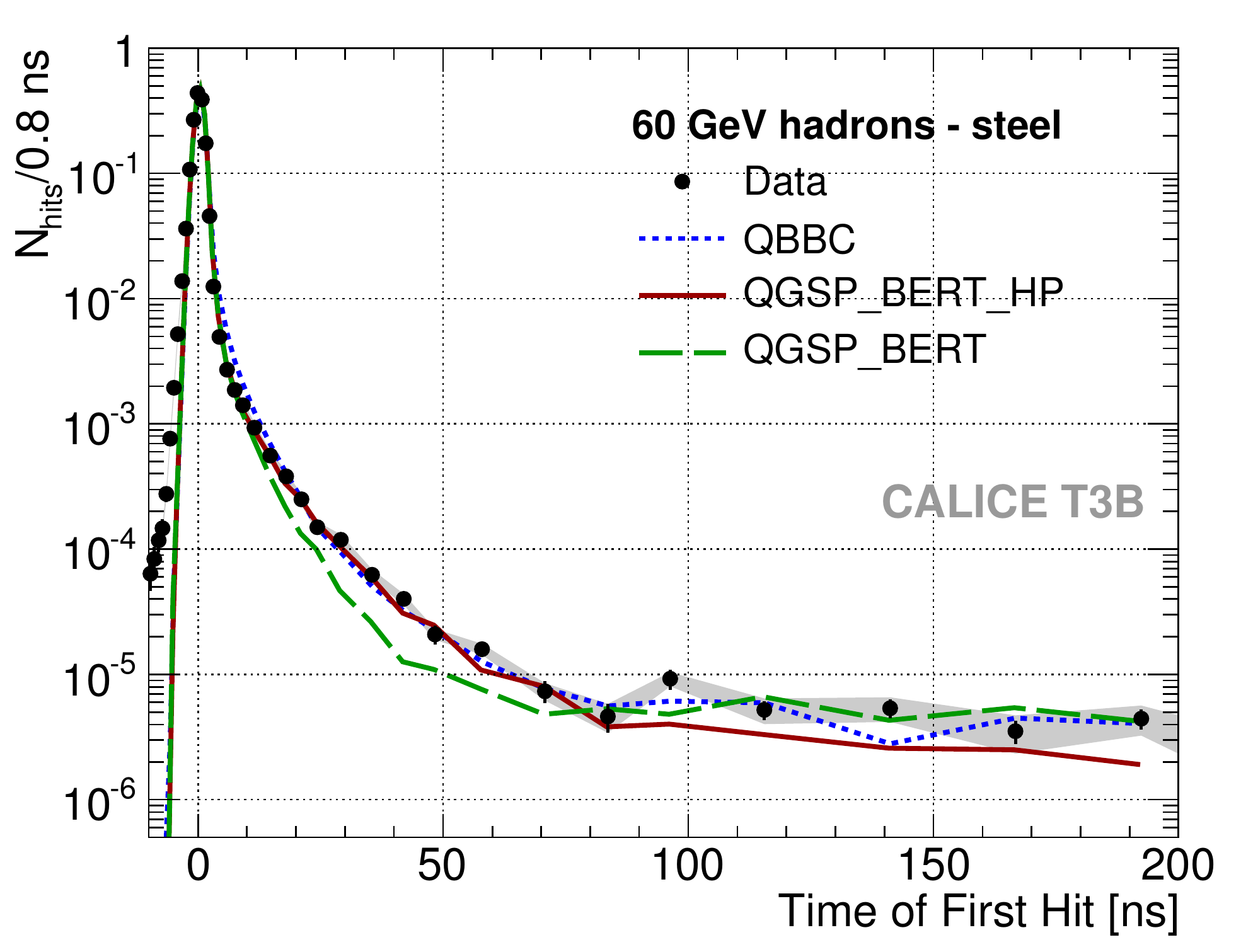} 
  \includegraphics[width=0.49\textwidth]{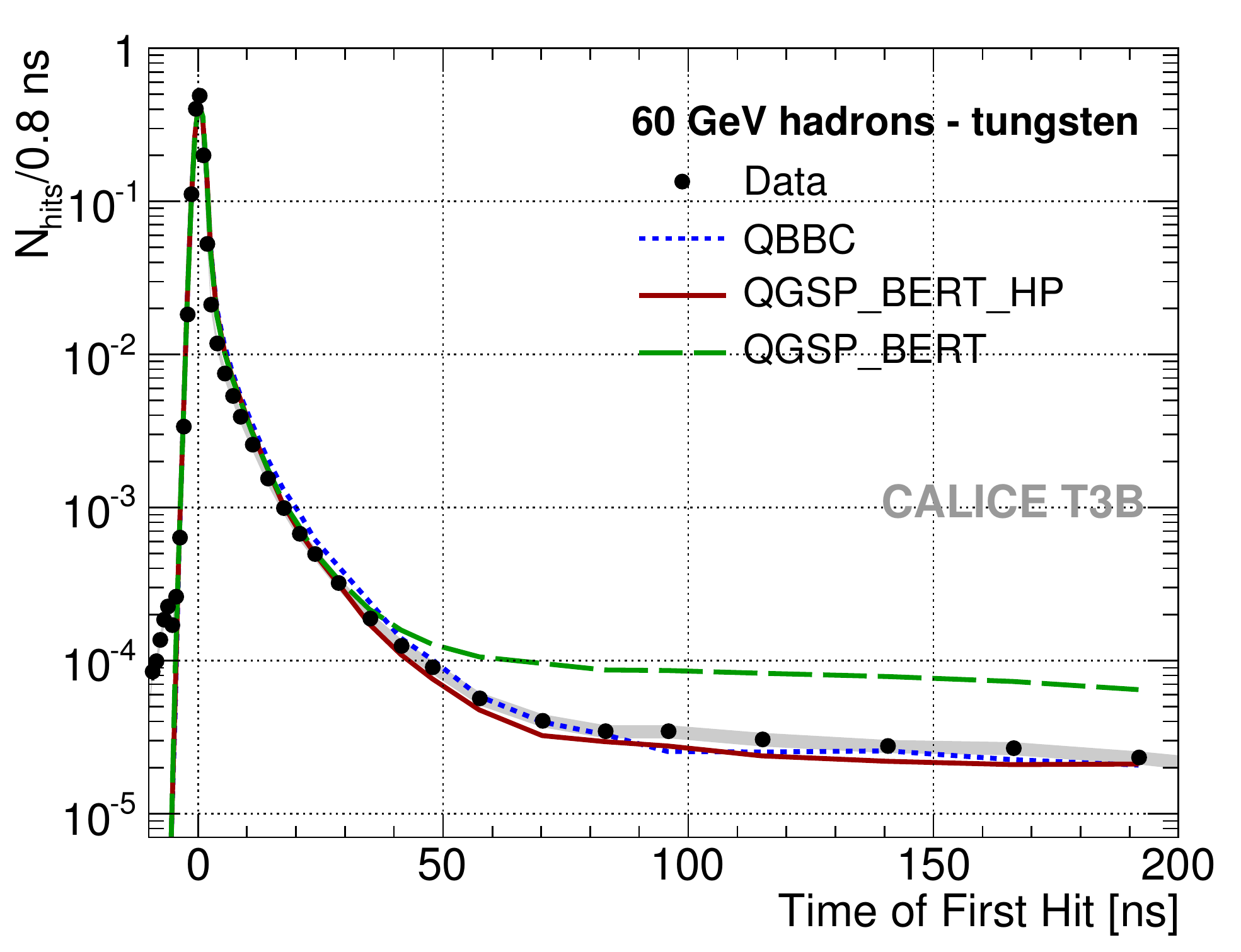}  

\caption{\label{fig:TungstenvsSteelSim}Distribution of the time of first hit in T3B for 60 GeV \textpi{}$^+$ in steel (left) and tungsten (right), compared to Geant4 simulations with different physics lists.}
\end{figure}

Finally, it is interesting to compare the prediction of simulations to experimental measurements. 
For the T3B data, detailed comparisons to GEANT4\cite{agostinelli2003geant4} simulations with different hadronic shower models have been performed for both absorbing materials. These show that while the distributions in steel are generally well modeled by all physics lists, the reproduction of the tungsten results requires a more precise neutron treatment. Indeed, as shown in figure \ref{fig:TungstenvsSteelSim}, physic lists like QGSP-BERT fail to reproduce data over time windows longer than a few tens of ns, whereas QGSP-BERT-HP, an extension of the former list with tracking of neutrons down to thermal energies, describe experimental results over a broader range. For the FastRPC experiment the simulation codes are still being developed at present.

\section*{References}
\bibliographystyle{iopart-num}
\bibliography{iopart-num}

\end{document}